\documentclass[a4,aps,twocolumn, showpacs,floatfix]{revtex4-1}
\usepackage{graphicx}
\usepackage{dcolumn}
\usepackage{color}
\usepackage{amsmath}
\usepackage{here}

\begin{document}

    \title{Nonlinear dynamics of Josephson vortices in merging superfluid rings}
\author{ Artem Oliinyk$^{1}$, Boris Malomed$^{2}$, and Alexander Yakimenko$%
^{1}$ }
\affiliation{$^1$ Department of Physics, Taras Shevchenko National University of Kyiv,
64/13, Volodymyrska Street, Kyiv 01601, Ukraine \\
$^2$ Department of Physical Electronics, Faculty of Engineering, and Center
for Light-Matter Interaction, Tel Aviv University, Tel Aviv 69978, Israel}

\begin{abstract}
We consider merger of two parallel toroidal atomic Bose-Einstein condensates
with different vorticities in a three-dimensional (3D) trap. In the
tunnel-coupling regime, Josephson vortices (rotational fluxons) emerge in
the barrier between the superflows. When the barrier is gradually
eliminated, we observe essentially three-dimensional evolution of quantum
vortices, which may include the development of the Kelvin-Helmholtz
instability at the interface between the rings, in the framework of a weakly
dissipative Gross-Pitaevskii equation. An initially more populated ring,
carrying a persistent current, can drag an initially non-rotating less
populated one into the same vortex state. The final state of the condensate
crucially depends on an initial population imbalance in the double-ring set,
as well as on the shape of the 3D trapping potential, oblate or prolate.
In the prolate (axially elongated) configuration, robust 3D \textit{hybrid
structures} may appear as a result of the merger of persistent currents
corresponding to different vorticities.
\end{abstract}

\maketitle

\section{Introduction}

One of the most remarkable manifestations of quantum properties of
superconductors \cite{Ustinov} and superfluids \cite{Kaurov2005, Kaurov2006}
is formation of Josephson vortices (JVs), alias fluxons, in long Josephson
junctions. In particular, much interest to JVs in atomic Bose-Einstein
condensates (BECs) has been drawn since circulating atomic supercurrents,
counterpropagating in two parallel 1D BEC, were theoretically investigated
in Refs. \cite{Kaurov2005, Kaurov2006}. Multidimensional JVs in
spin-orbit-coupled BECs were considered in Ref. \cite{PhysRevA.93.033618}.
Rotational fluxons in BEC loaded into coplanar double-ring traps were
studied in Ref. \cite{Brand2009}. Further, it was demonstrated in Ref. \cite%
{BrandPRL13} that atomic BECs, confined by a dual-ring trap, support JVs as
topologically stable defects in the filed of the relative phase. Stationary
JVs were predicted in an array of linearly-coupled one-dimensional
Bose-Einstein condensate in Ref. \cite{Brand18}. Very recently, JVs were
investigated in a closed-loop stack of 1D and 2D atomic BECs \cite%
{MateoNJP2019}.

In our recent work \cite{arxiv19} it was demonstrated that the azimuthal
structure of the tunneling flow implies formation of Josephson vortices
(rotational fluxons) with zero net (integral) current through the junction
for hybrid states, built of persistent currents with different topological
charges in coupled rings (in particular, these include the case of opposite
topological charges -- this is the configuration which was defined as hybrid
states in Ref. \cite{peanut}, where they were supported by an effective
nonlinear potential). It turns out that the fluxons' cores rotate and bend,
following the action of the quench, i.e. formation of tunnel junction with chemical potential difference.
Further, it was
found that, as the barrier decreases, and the effective coupling between the
rings respectively increases, the JVs accumulate more and more energy. This
scenario suggests the question if all quantum vortices, which are present in
the settings, \textit{viz}., the fluxons and circular currents, are always
unstable, and what the final state of the merging counter-propagating
superflows is in the double-ring system.

In this connection, it is relevant to mention that the Kelvin-Helmholtz
instability (KHI), which is well known in classical hydrodynamics, leads to
growth of roll-up perturbations at the interface of two streams with
different velocities \cite{Peltier}. KHI plays a crucial role in destroying
laminar flows, driving them into a turbulent state. In classical viscous
liquids and gases, adjacent streams with initially different velocities
evolve into a state with the averaged velocity. For superfluids the dynamics
of KHI is fundamentally different \cite{Tsubota13}. In particular, a pair of
weakly interacting persistent currents in toroidal Bose-Einstein condensates
(BECs), carrying quantum vortices, may undergo transitions between different
angular-momentum states via phase slips, at which the angular momentum
changes by unitary quanta. In sharp contrast to classical streams, there is
no state with an \textquotedblleft intermediate velocity" for two
interacting persistent currents initially created with different
vorticities. The quantum nature of\ the transition between different
angular-momentum states leads to far-reaching consequences for the
development or suppression of the KHI in ultracold atomic gases.

Dynamical KHI at the interface in phase-separated two-component
Bose-Einstein condensates was investigated in Refs. \cite%
{Tsubota2010,Tsubota2010_2}. In recent theoretical work \cite%
{PhysRevA.97.053608} two merging counter-propagating streams have been
investigated in the framework of conservative 2D (two-dimensional)
Gross-Pitaevskii equation (GPE). Planar counterflow channels with periodic
boundary conditions mimic two narrow (in the radial direction) ring-shaped
condensates which merge when the separating barrier is switched off.
However, the description of KHI in realistic atomic BECs calls for a more
detailed analysis. On the one hand, vortex lines, which drive KHI in the
double-ring system, are oriented along the orthogonal axis, which requires
to introduce a full 3D model for the consideration of the vortex dynamics.
Further, weak dissipative effects must be taken into account to analyze the
relaxation to the stationary state after the transition induced by KHI.

In the present work we address the following issues: (i) How nonlinear dynamics of
quantum vortices drives the relaxation of two merging persistent currents,
and (ii) identification of a final state of two merging ring-shaped
condensates with different initial vorticities. The analysis reveals two
noteworthy effects: (i) In pancake-shaped configurations (those compressed
in the axial direction), a more populated ring can \emph{impose its quantum
state} onto the originally less populated one. In particular, if the ring
carries topological charge $1$, the persistent current merges with a
non-rotating ring. The final state of the resulting toroidal condensate is
either one with zero angular momentum, or a single-charged quantized flow,
depending on the initial difference in populations of the two rings. (ii)
Instead of the development of KHI at the interface of the merging persistent
currents, we observe formation of \emph{robust hybrid vortex structures} in
the potential trap sufficiently elongated in the axial direction.

The rest of the paper is organized as follows. The model is formulated in
Section II. Results of the systematic numerical analysis are summarized in
Section III, separately for the pancake-shaped and elongated shapes of the
trapping potential. The paper is concluded by Section IV.


\section{The model}


In modeling nonequilibrium dynamics, such as quantum turbulence \cite%
{Tsubota13} or nucleation of vortices \cite{PRA13}, dissipative effects,
even if they are weak, are of crucial importance for relaxation to
equilibrium states. In particular, the dissipation drives the drift of the
vortex core to the edge of the BEC cloud \cite{PhysRevA.81.023630,2012PhRvA..86a3629M, PRA2013,PhysRevA.87.013630, PhysRevA.92.053603}.
Such effects naturally arise in
a trapped condensate due to its interaction with a thermal component, and
can be captured phenomenologically by the dissipative GPE derived by Choi
\textit{et al}. \cite{Choi,Proukakis}. Close to the thermodynamic
equilibrium, the weakly dissipative GPE is written as
\begin{equation}
(i-\gamma )\hbar \frac{\partial \psi }{\partial t}=-\frac{\hbar ^{2}}{2M}%
\nabla ^{2}\psi +V_{\text{ext}}(\mathbf{r},t)\psi +g|\Psi |^{2}\psi -\mu
\psi ,  \label{GPE3D}
\end{equation}%
where $g=4\pi a_{s}\hbar ^{2}/M$ is the coupling strength, $M$ is the atomic
mass ($M=3.819\times 10^{-26}$ kg for $^{23}$Na atoms), $a_{s}$ is the $s$%
-wave scattering length (positive $a_{s}=2.75$ nm, corresponding to the
self-repulsion in the same atomic species, is used below), $\mu $ is the
chemical potential of the equilibrium state, and $\gamma \ll 1$ is a
phenomenological dissipative parameter. This form of the dissipative GPE has
been used extensively in previous studies of vortex dynamics (see, e.g.,
\cite{Tsubota13, PRA.77.023605, PRA.67.033610, PRA13}). In what follows
below, we assume $\gamma $ to be spatially uniform, and set $\gamma =0.03$
as in Refs. \cite{Tsubota13, PRA2013}. Actually, we have verified that
results reported below do not essentially depend on a specific value of $%
\gamma \ll 1$.

We consider a toroidal condensate, split by a blue-detuned sheet beam in
upper and lower weakly coupled rings-shaped components. The respective total
trapping potential is
\begin{equation}
V_{\mathrm{ext}}(\rho ,z,t)=\frac{1}{2}M\omega _{r}^{2}(\rho -\rho _{0})^{2}+%
\frac{1}{2}M\omega _{z}^{2}z^{2}+V_{\text{b}}(z,t),  \label{24}
\end{equation}%
where $\rho \equiv \sqrt{x^{2}+y^{2}}$ and the sheet potential is
\begin{equation}
V_{\text{b}}(z,t)=U_{\text{b}}(t)\exp \left( -\frac{1}{2}\frac{(z-z_{0})^{2}%
}{a^{2}}\right) ,  \label{Vb}
\end{equation}%
with the time-dependent strength,%
\begin{equation}
U_{\text{b}}(t)=\left\{
\begin{array}{c}
(1-t/t_{d})u_{b},~\text{at~}~t<t_{d}, \\
0,~~\text{at~}~t>t_{d}.%
\end{array}%
\right.  \label{Ub}
\end{equation}
Here the switching time is $t_{d}=0.015$ s, and $z_{0}$ is a possible shift
of the barrier shift along the $z$-axis.

For numerical simulations of the 3D GPE we use scaled time, $t\rightarrow
t\omega _{r}$, length $\mathbf{r}\rightarrow \mathbf{r}/l_{r}$, chemical
potential $\mu \rightarrow \mu /(\hbar \omega _{r})$, external potential $V_{%
\text{ext}}\rightarrow V_{\text{ext}}/(\hbar \omega _{r})$, and wave
function $\psi \rightarrow \psi \cdot l_{r}^{3/2}$, which casts GPE (\ref{GPE3D})
in the following form:
\begin{equation}
(i-\gamma )\frac{\partial \psi }{\partial t}=-\frac{1}{2}\nabla ^{2}\psi +{V}%
_{\text{ext}}\psi -\mu \psi +g|\psi |^{2}\psi ,  \label{GPE_dimless}
\end{equation}%
where the scaled positive nonlinearity strength is $g=4\pi a_{s}/l_{r}$, and
the scaled trapping potential is
\begin{equation}
V_{\text{ext}}=\frac{1}{2}(\rho -\rho _{0})^{2}+\frac{1}{2}A^{2}z^{2}+V_{%
\text{b}},  \label{potential}
\end{equation}%
where the aspect ratio of the toroidal trap is
\begin{equation}
A=\omega _{z}/\omega _{r}~.  \label{A}
\end{equation}%
It turns out that dynamics of the quantum vortices, observed after the
merger of the rings, crucially depends on $A$. In this work we first
concentrate on the pancake-shaped trapping potential with typical values of
the trapping frequencies \cite{Wright13,FredPRL14}: $\omega _{r}=2\pi \times
123$ Hz and $\omega _{z}=2\pi \times 600$ Hz, hence $A=4.88$, the oscillator
length of the longitudinal trapping potential is $l_{r}=\sqrt{\hbar /\left(
M\omega _{r}\right) }=1.84$ $\mathrm{\mu }$m, $\rho _{0}=19.23$ $\mathrm{\mu
}$m, and $g=1.88\cdot 10^{-2}$. Subsequently, we address the evolution of
the merging rings for an elongated trap with $A<1$. Scaled parameters of the
potentials in Eqs. (\ref{Vb}) and (\ref{Ub}) are fixed to be $a=0.3,u_{b}=80$%
. 
In the text following below, we use the same notation for the scaled wave
function $\psi $, spatial coordinates $\left( x,y,z\right) $, and time $t$
as above, as it will produce no confusion.


A steady-state solution of Eq. (\ref{GPE_dimless}) was found numerically by
dint of the imaginary-time-propagation method, applied with $\gamma =0$. To
obtain stationary rings, with different vortex phase profiles in the upper
and lower rings, the evolution in imaginary time was initiated by the
following ansatz:
\begin{equation}
\Psi (\mathbf{r})=|\Psi _{0}(x,y,z)|e^{iS(z)\phi },  \label{input}
\end{equation}%
where $\phi $ is the polar angle in cylindrical coordinates, and different
integer topological charges $m_{1}$ and $m_{2}$ are imprinted in the top and
bottom rings: $S(z)=m_{1}$ for $z<z_{0}$ and $S(z)=m_{2}$ for $z\geq z_{0}$,
cf. a similar procedure used in the model with a \textquotedblleft
peanut-shaped" nonlinear trapping potential in Ref. \cite{peanut}. The
imaginary-time propagation converges to steady states with required accuracy
for an arbitrary input amplitude $|\Psi _{0}(x,y,z)|$ in Eq. (\ref{input})
with a fixed norm:
\begin{equation}
\langle \Psi _{0}|\Psi _{0}\rangle =N\equiv N_{1}+N_{2},  \label{N}
\end{equation}%
where $N_{1}$ and $N_{2}$ are scaled numbers of atoms in the bottom and top 
rings, respectively:
\begin{eqnarray}
N_{1} &=&\int_{-\infty }^{+\infty }dx\int_{-\infty }^{+\infty
}dy\int_{-\infty }^{z_{0}}|\Psi _{0}(x,y,z)|^{2}dz,  \notag \\
N_{2} &=&\int_{-\infty }^{+\infty }dx\int_{-\infty }^{+\infty
}dy\int_{z_{0}}^{+\infty }|\Psi _{0}(x,y,z)|^{2}dz  \label{12}
\end{eqnarray}%
.

Note that, by shifting center $z_{0}$ of the splitting barrier (\ref{Vb}),
it is easy to prepare an initial state with dominant population in the ring
with topological charge $m_{1}$ ($N_{1}>N_{2}$ for $z_{0}>0$) or $m_{2}$ ($%
N_{2}>N_{1}$ for $z_{0}<0$), the respective asymmetry parameter being
\begin{equation}
P=(N_{1}-N_{2})/(N_{1}+N_{2}).  \label{P}
\end{equation}%
%
%

It is relevant to mention two-component models with incoherent nonlinear
interaction between the components, which conserve the norms in each
component separately. Unlike the present setting, such systems readily admit
stationary states with different vorticities and different chemical
potentials in the components. In particular, systems of this type give rise
to stable states with \textquotedblleft hidden vorticity", i.e., ones with
opposite vorticities and equal norms in the two components, the total
angular moment of the states being zero, as predicted in BEC \cite%
{hidden1,hidden4,hidden6,hidden7,hidden8,hidden10,hidden11,
hidden13,hidden14} and optics \cite{hidden5,hidden9,hidden12}.


\section{The evolution of merging persistent currents}


The dynamics of merging toroidal condensates in real time was simulated by
means of the split-step fast-Fourier-transform method. The barrier strength $%
U_{\text{b}}(t)$ in potential (\ref{24}) decreasing in time until extinction
at $t=t_{d}$, two rings tend to merge into a single toroidal condensate at $%
t>t_{d}$.

We considered different regimes of the dissipative evolution, controlled by
parameter $\gamma $ in Eq. (\ref{GPE3D}). In the conservative limit, with $%
\gamma =0$ and a constant trapping potential, the total norm, which is
defined by Eqs. (\ref{N}) and (\ref{12}), energy
\begin{equation}
E=\int \left[ \frac{1}{2}|\nabla \psi |^{2}+V_{\text{ext}}(\mathbf{r})|\psi
|^{2}+\frac{g}{2}|\psi |^{4}\right] d\mathbf{r},  \label{E}
\end{equation}%
and the $z$-component of the total angular momentum,%
\begin{equation}
\mathbf{L}=-\frac{i}{2}\int \left\{ \psi ^{\ast }\left[ \mathbf{r}\times
\nabla \psi \right] -\psi \left[ \mathbf{r}\times \nabla \psi ^{\ast }\right]
\right\} d\mathbf{r,}  \label{L}
\end{equation}%
are conserved in our simulations with high accuracy. As seen in Fig. \ref%
{E_vs_t} in non-dissipative case ($\gamma =0$), the total energy decays at $%
0<t<t_{d}$, whilst the potential barrier between the two rings gradually
vanishes, and remains constant at $t>t_{d}$.

The system described by Eq. (\ref{GPE_dimless}) with $\gamma>0$
conserves neither the energy nor the number of particles. The chemical
potential $\mu(t)$ of the equilibrium state in our dynamical simulations was
adjusted at each time step so that the number of condensed particles slowly
decays with time: $N(t) = N(0)$exp$(-t/t_0)$ , where $t_0 = 10$ s
corresponds to the $1/e$ lifetime of the BEC reported in the experiment \cite%
{Hadzibabic}. In our simulations we assume that total number of atoms in the
initial state is $N(0)=6\cdot 10^{5}$.

The goal of the numerical analysis is two-fold. First, we aim to trace the
3D dynamics of quantum vortices, in the course of the relaxation following
the vanishing of the separating barrier. The second objective is to identify
final states of the dissipative evolution for different values of the
initial asymmetry parameter (\ref{P}).

%
\begin{figure}[h]
\centering
\includegraphics[width=3.4in]{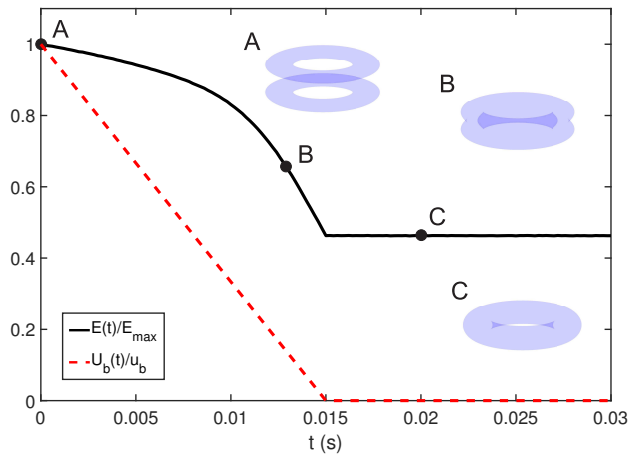}
\caption{(Color online) The ratio of the total energy [defined as per Eq. (%
\protect\ref{E})] to the initial value $E(t)/E(0)$ (the black solid curve)
for the merging rings with vorticities $(0,0)$ vs. time in the
dissipationless system ($\protect\gamma =0$,  $\omega_z/\omega_r=4.88$). The decay of the energy at $%
t<t_{d}$ follows the gradual vanishing of the potential barrier separating
the two ring-shaped condensates, as per Eq. (\protect\ref{Ub}). The
corresponding normalized $U_{\text{b}}(t)/U_{\text{b}}(0)$ barrier's
strength is shown by the red dashed line. Insets: examples of 3D isosurfaces
of the constant condensate density at moments of time corresponding to
points $A,B$, and $C$ marked at the energy curve. }
\label{E_vs_t}
\end{figure}
\begin{figure}[h]
\centering
\includegraphics[width=3.4in]{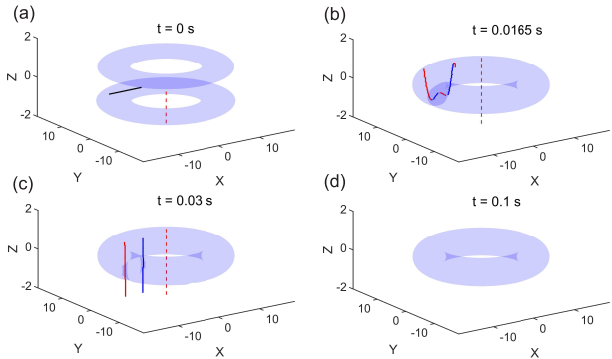}
\caption{(Color online) The evolution of the merging rings when the
population in the bottom ring with $m_{1}=1$ slightly dominates over the top
one, with $m_{2}=0$ [imbalance parameter (\protect\ref{P}) is $P=0.06$, the
dissipation parameter in Eq. (\protect\ref{GPE_dimless}) being $\protect%
\gamma =3\cdot 10^{-2}$]. Shown are snapshots of 3D isosurfaces of constant
condensate density. The axis of the JV (Josephson vortex) at $t=0$, which
exists due to the structure of the supercurrent tunneling between the rings
with different vorticities (see the text), is indicated by the black line.
The vortex' axis trapped in the internal toroidal hole is marked by the
dashed red line. While the rings are merging, the fluxon bends and splits
into a vertically oriented vortex (the red curve, drifting to the the
external periphery) and an anti-vortex (the blue curve), drifting towards
the central toroidal hole. The antivortex annihilates with the central
vortex line, which leads to a final state with zero total angular momentum
(see also Fig. \protect\ref{2Lp}). Full evolution one can see in the supplemental movie \cite{sup_mat}.}
\label{CoresV0}
\end{figure}
\begin{figure}[h]
\centering
\includegraphics[width=3.4in]{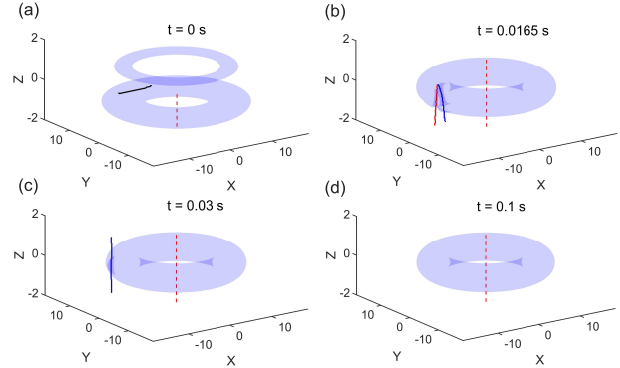}
\caption{(Color online) The same as in Fig. \protect\ref{CoresV0} for the
case when the population in the bottom ring with $m_{1}=1$ strongly
dominates in the initial state ($P=0.49$). In this case, both vortex and
antivortex lines, produced by the splitting of the bending fluxon, escape
from the condensate through the peripheral region. Eventually, the merger of
the rings leads to the emergence of a single-charged ($m=1$) persistent
current, see also Fig. \protect\ref{2Lp}. Full evolution one can see in the supplemental movie \cite{sup_mat}.}
\label{CoresV1}
\end{figure}
\begin{figure}[h]
\centering
\includegraphics[width=3.4in]{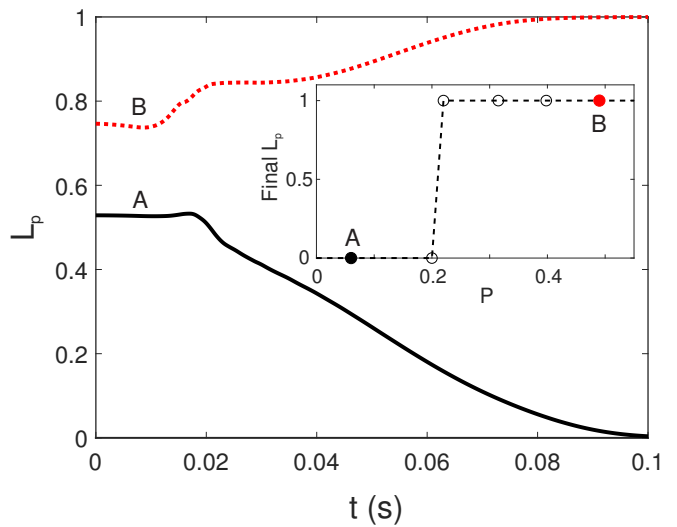}
\caption{(Color online) The evolution of the total angular momentum per
particle, $L_{p}=L_{z}/N$, for the states with initial vorticities $%
(m_{1}=1,m_{2}=0)$ and different values of the initial population imbalance,
defined as per Eq. (\protect\ref{P}): A) $P=0.06$; B) $P=0.49$, the
dissipation parameter being $\protect\gamma =3\cdot 10^{-2}$. The inset
shows the final value of $L_{p}$ as the function of the initial imbalance, $P
$. The dominant ring, which carries a persistent current, drags the
initially non-vortical one into the same vortex state, with $m=1$, at $P>P_{%
\text{cr}}\approx 0.21$.}
\label{2Lp}
\end{figure}
\begin{figure}[h]
\centering
\includegraphics[width=3.4in]{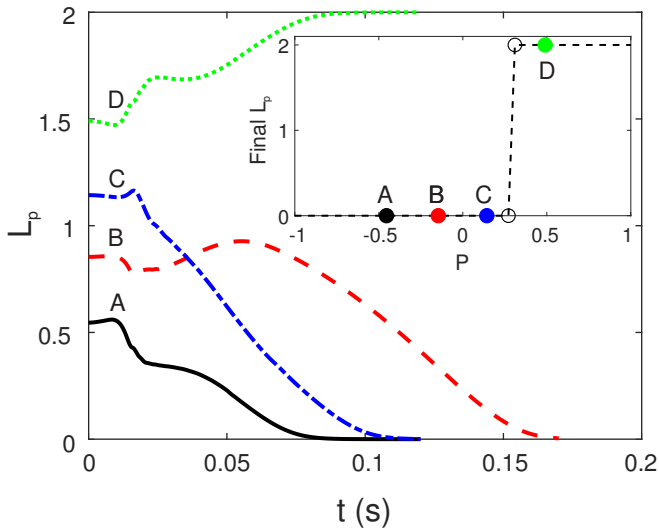}
\caption{(Color online) Evolution of the total angular momentum per particle
$L_{p}=L_{z}/N$ for inputs with initial vorticities $(m_{1}=2,m_{2}=0)$ and
different values of the initial population imbalance, defined as per Eq. (%
\protect\ref{P}): A) $P=-0.45$; B) $P=-0.15$; C) $P=0.14$; D) $P=0.49$, the
dissipation parameter being $\protect\gamma =3\cdot 10^{-2}$. The inset
shows the final value of $L_{p}$ as a function of the initial imbalance, $P$%
. At $P>P_{\text{cr}}\approx 0.29$ the system relaxes to the state with the
double-charged persistent current ($m=2$), while a final state with $m=1$ ($%
L_{p}=1$) is not observed.}
\label{4Lp}
\end{figure}

\begin{figure*}[tbh]
\centering
\includegraphics[width=6.8in]{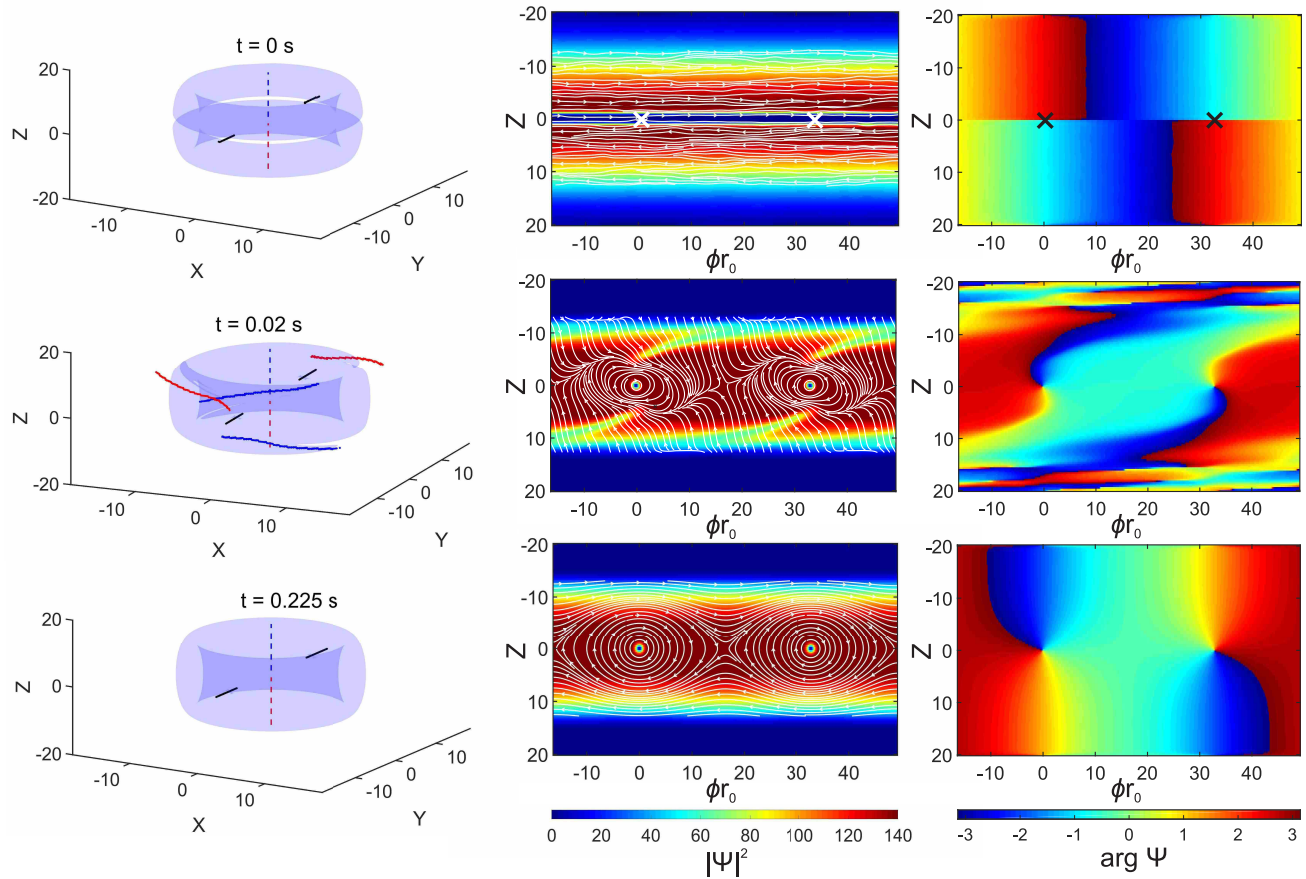}
\caption{(Color online) Snapshots at three moments of time, illustrating the
evolution of the merging strongly elongated toroidal condensates with a
small aspect ratio, $A=0.16$ [see Eq. (\protect\ref{A})] and vorticities $%
m_{1,2}=\pm 1$ in its two axially separated parts. The input has no
imbalance between the parts ($P=0$), the dissipative constant in Eq. (%
\protect\ref{GPE_dimless}) being $\protect\gamma =0.03$. Shown are 3D
isosurfaces with constant condensate density (left), and maps of the
distribution of the density (middle) and phase (right) on the cylindrical
surface of radius $\protect\rho _{0}$ [see Eq. (\protect\ref{potential})], $%
\protect\phi $ being the angular coordinate. A long-lived hybrid complex is
produced by the evolution.}
\label{fig:Hybrids}
\end{figure*}
\begin{figure}[h]
\centering
\includegraphics[width=3.4in]{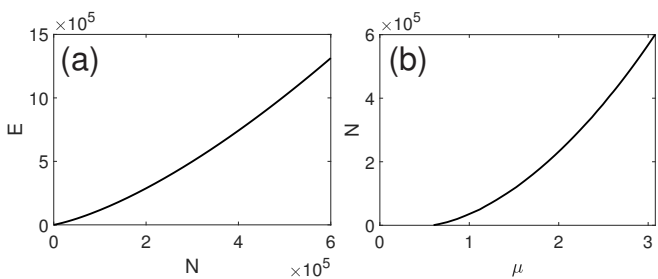}
\caption{(a) Total energy of the hybrid vs. number of particles for $m_{1}=1$%
, $m_{2}=-1$, $P=0$, $A=0.16$; (b) Total number of particles of the
symmetric ($P=0$) hybrid vortex structure $m_{1}=1$, $m_{2}=-1$ vs. the
chemical potential for elongated condensate $A=0.16$.}
\label{fig:E_N_mu}
\end{figure}
\begin{figure*}[tbh]
\centering
\includegraphics[width=6.8in]{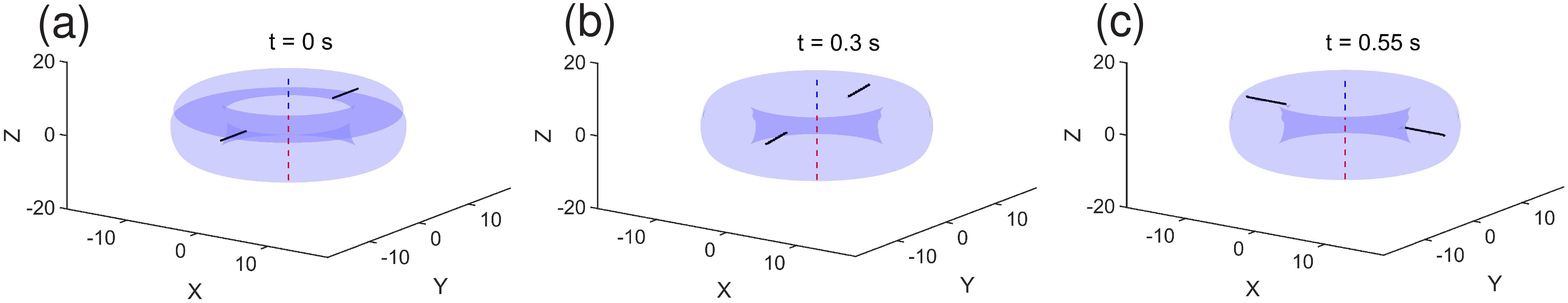}
\caption{Example of stable evolution of the vortex hybrid. The remaining
blue-detuned sheet beam not only stabilizes the vertical drift of the
fluxon's cores (shown by horisontal black lines) but also provides a
possibility for controllable manipulation of the total angular momentum of
the hybrid. The initial biased double-ring system $(+1,-1)$ separated by the
barrier with angular momentum per particle (a) $L_{p}=0.59$ merges into the
stable hybrid with pinned fluxons. Then for $t>0.3$ s $L_{p}$ gradually
decreases as position of the sheet beam is lowered: (b) $L_{p}=$0.58; (c) $%
L_{p}=$0.46}
\label{fig:FixCore}
\end{figure*}

\subsection{Pancake-shaped toroidal condensates}


Figures \ref{CoresV0} and \ref{CoresV1} provide typical examples of merging
pancake-shaped toroidal condensates [with aspect ratio $A=4.88$, see Eq. (%
\ref{A})], in the cases when the initial atomic population in the bottom
ring, with vorticity $m_{1}=1$, is, respectively, weakly or strongly
dominant [$P=0.06$ and $P=0.49$, see Eq. (\ref{P})] over the top one, with $%
m=0$.
The lower ring, carrying the persistent current in the initial state,
contains a single-charged vertical vortex line threading the internal
toroidal hole [the red dashed lines in Fig. \ref{CoresV0} (a) and \ref%
{CoresV1}] \cite{phase_detect}.

The fact that the initial states in the bottom and top rings have different
vorticities, $m_{1}\neq m_{2}$, implies that the superflow tunneling across
the separating potential creates $|m_{1}-m_{2}|$ JV lines (rotational
fluxons) located in the low-density region between the weakly interacting
rings \cite{arxiv19}. In the course of the evolution, the JV, which is
initially oriented horizontally between the rings, splits in vortices
oriented vertically along the $z$ axis, see Fig. \ref{CoresV0}(b,c) and \ref%
{CoresV1}(b,c)].
Thus, Figs. \ref{CoresV0} and \ref{CoresV1} demonstrate that the initial
states, with the horizontally-oriented JV and vertically-oriented vortex
line, evolve into series of vortices and antivortices, with axes oriented
parallel to the $z$-axis. Note that a similar snake instability and decay of
the JV into vortex dipoles was also observed in spin-orbit-coupled BECs \cite%
{PhysRevA.93.033618}. Subsequent evolution of the vortices depends on two
factors: (i) interaction between adjacent vortices (attraction for a
vortex-vortex pair and repulsion for an vortex-antivortex one), and (ii)
drift of the vortex line caused by the dissipation. Under the action of the
weak dissipation, the vortex lines spiral out from the bulk of the atomic
cloud towards low-density regions at the surface of the condensate. This may
imply slow drift of the vortex line either to the external surface of
the toroidal condensate, where the vortex decays into elementary
excitations, or trapping of the vortex line in the central toroidal hole,
which adds a positive or negative unit to the topological charge of the
persistent current.


One can see from the Fig. \ref{2Lp}, that for the initial state with \emph{%
weak asymmetry} ($P=0.06$, $z_{0}=0.02$), the evolution of which is
displayed in Fig. \ref{CoresV0}, the angular momentum per particle, $%
L_{p}=L_{z}/N$, vanishes as a result of the relaxation [the total angular
moment is defined in Eq. (\ref{L})]. This observation agrees with that outlined in the caption to Fig. \ref{CoresV0}, \textit{viz}.,
that, under the action of the dissipation, the vortex line drifts in Figs. %
\ref{CoresV0}(c,d) to the peripheral (external) low-density area and decays,
while the antivortex drifts towards the central $z$ axis and eventually
annihilates with the central vortex (shown by dashed red line). In this case
the system relaxes to the irrotational ground state, as seen in Fig. \ref%
{CoresV0}(d).

In the case of \emph{strong asymmetry} ($P=0.49$, $z_{0}=0.175$) the value
of $L_{p}$ evolves towards $L_{p}=1$, see Fig. \ref{2Lp} (b). As seen in
Fig. \ref{CoresV1}(c,d) and is outlined in the caption to the figure, the
fluxon bends and splits into a vortex-antivortex dipole, which escape from
the condensate. The final state contains the central vortex (designated by
the red dashed line) corresponding to the single-charged ($m=1$) persistent
current, as seen in Fig. \ref{CoresV1}(d).

Systematic simulations demonstrate that there is a threshold value of the
initial asymmetry parameter, $P_{\text{cr}}$ [defined as per Eq. (\ref{P})],
in the pancake-shaped double-ring system, which determines the final state
of the toroidal condensate after the relaxation (see the inset in Fig. \ref%
{2Lp}). After long-time dissipative evolution, the toroidal condensate
relaxes to the state with zero angular momentum at $P<P_{\text{cr}}$,
while at $P>P_{\text{cr}}$ the dominant component (with $m_{1}=1$) imposes
its quantum state onto the less populated ring (with $m_{2}=0$), so that the
entire condensate relaxes to the state with $m=1$. Summarizing results of
the simulations, we conclude that the critical initial asymmetry takes
values in interval $0.20<P_{\text{cr}}<0.22$.

The conclusion that the more populated ring, carrying the persistent current
with $m=1$, can drag an initially zero-vorticity one into the same vortex
state with $m=1$ does not contradict the energy balance (in the
dissipationless system, with $\gamma =0$). The potential energy of the
condensate, trapped by oscillatory potential in $z$-direction, decreases
when the horizontal repulsive separating barrier is eliminated. Thus the
total energy of the initial double-ring system is greater than the energy of
the final single-ring condensate state, both for $L_{p}=0$ and $L_{p}=1$.
As is known, in toroidal condensates a persistent current with topological
charge $0\leq m\leq m_{\text{max}}$ corresponds to the stable state with a
local energy minimum, where the maximum topological charge $m_{\text{max}}$
admitting the stability depends on the condensate parameters (see, e.g.,
\cite{2012PhRvA..86a3629M,PRA2015}). 
Thus, by reducing the potential barrier between the rings with charges $%
m_{1}=1$ and $m_{2}=0$, one triggers the relaxation of the appearing single
toroidal condensate either to the zero-vorticity ($L_{p}=0$) ground state,
or to the stable state with $L_{p}=1$, depending on the initial value of
imbalance $P$, defined as per Eq. (\ref{P}).

We have performed a similar analysis for the double-ring system with initial
topological charges $(m_{1}=2,m_{2}=0)$. Figure \ref{4Lp} displays the
evolution of the angular momentum per particle for different initial values
of the imbalance, $P$, in this case. Note that, at $P>P_{\text{cr}}\approx
0.29$, the system relaxes to the state with double-charged ($m=2$)
persistent current. Note that the final state with $m=1$ is not observed in
the present case, even for the symmetric input with $P=0$ and $L_{p}=1$,
which is closest to a possible outcome with $m=1$. The point is that, due to
the symmetry of the system, two fluxons featured by the initial
configuration split into vertical vortices which either escape towards the
external periphery (which leads to the final state with $m=0$), or both
vortex cores are trapped together in the central toroidal hole, which
corresponds to the final state with $m=2$. Detailed consideration of
spontaneous symmetry breaking in the course of the relaxation process after
the merger of superflows with different vorticities may be a relevant
extension of the present work.

It is interesting to compare our results for the 3D merging rings (where the
quantized vortices drive the relaxation dynamics) and
the merger of two concentric 2D BECs, studied in Refs.
\cite{Tsubota2018,Tsubota2019}, which reported the emergence of a spiral
dark soliton when one condensate carried a nonzero initial angular momentum.
The spiral dark soliton enables the transfer of the angular momentum between
the condensates, allowing the merged condensate to rotate even in the
absence of the overall vorticity. 

\subsection{Elongated toroidal condensates}

Our next objective is to compare the evolution of the rings elongated in the
vertical ($z$) direction, with small values of aspect ratio (\ref{A}), and
those supported by the pancake-shaped toroidal traps with large values of $%
A,$ which were considered above. Thus, we aim to figure out
the role of the trapping-potential geometry in the JV\ dynamics.

Note that a strongly elongated condensate is related to the 2D setup
considered in Ref. \cite{PhysRevA.97.053608}, thus one may expect formation
of a vortex lattice similar to
one observed in the 2D single-component setting \cite%
{Sakaguchi2010,PhysRevA.97.053608}, or a vortex sheet observed in
two-component \cite{TsubotaSheets2009} condensates. The lattice of 2D
vortices in the 3D setup corresponds to fluxons at an interface of
superflows with different vorticities.

The simulations clearly demonstrate that 3D hybrid states, built of two
axially separated parts with different vorticities, emerge as a result of
the evolution of the merging elongated rings with different vorticities. A
typical example is displayed in Fig. \ref{fig:Hybrids}, for unequal
vorticities $m_{1,2}=\pm 1$, aspect ration $A=0.16$ [see Eq. (\ref{A})], and
zero initial imbalance between the two parts, $P=0$ [see Eq. (\ref{P})].
Strictly speaking, the emerging hybrid states are transient ones, as they
will eventually transform into usual ones. However, they remain robust in
the course of long evolution, and do not decay even being strongly
perturbed; instead, perturbed hybrids relax back to the unperturbed shape.
velocity of
the vertical drift of the fluxons is determined by the dissipation rate
parameter $\gamma $ and gradient of the condensate in the $z$-direction: the
greater the dissipation, the faster fluxons drift to the region with lower
condensate density. Since the condensate's density distribution is nearly
uniform halfway between two elongated condensates, which inhibits
degradation of the hybrid.
A typical example of hybrid decay is shown in in Supplemental
video \cite{sup_mat}, where the lifetime of hybrid with particles imbalance $P=0.1$ is about $0.7 s$.
The lifetime of the hybrids in our simulations depends on the imbalance $P$ and for the symmetric setup $P\approx 0$ a perturbed hybrid can survive during several seconds, which is comparable with duration of typical experiment.

Families
of the long-lived hybrids may be well characterized by the respective
dependencies of the total number of particles on the chemical potential, and
of the total energy on the number of particles, as shown in Fig. \ref%
{fig:E_N_mu}.

These findings are in sharp contrast with the results of 2D simulations
reported in Ref. \cite{PhysRevA.97.053608}, where unstable evolution of the
vortex street led to the onset of the KHI-driven dynamics. Note that the
decay of the vortex lattice and development of KHI was demonstrated in Ref.
\cite{PhysRevA.97.053608} for $|m_{1}-m_{2}|=20$, hence the respective
vortex street contained $40$ vortex cores. As pointed out in \cite%
{PhysRevA.97.053608}, in the absence of perturbations the vortex streets
survived for a long time; however, under the action of noisy perturbations
(whose magnitude was less than one percent of the background wave function)
the 2D vortex lattice rolled-up, which was interpreted in as the development
of KHI \cite{PhysRevA.97.053608}.

Thus a natural question arises if KHI actually develops in full 3D system
for $|m_{1}-m_{2}|~\sim 1$. In our simulations, the lattice of Josephson
vortices is not affected by KHI even in the case of relatively strong noisy
perturbations, with amplitudes up to $\sim 5\%$ of the unperturbed state.

As shown in Fig. \ref{fig:Hybrids}, following the merger of the rings with
topological charges $m_{1}=1$ and $m_{2}=-1$, vortex lines are horizontally
oriented, in sharp contrast to the vertically oriented vortices observed in
Figs. \ref{CoresV0} and \ref{CoresV1}. Results of our 3D simulation are in qualitative agreement with findings
reported in Ref. \cite{PhysRevA.97.053608}, which were obtained in the
framework of the simplified 2D model for merging flows with vorticities $%
m_{1}=+1$ and $m_{2}=-1$.

In our simulations the development of the snake-type instability is strongly
suppressed for elongated condensates with aspect ratio $A<1$, see Eq. (\ref%
{A}). Further, it was found that for the strongly elongated toroidal traps
with $A<0.6$ JVs keep the radial orientation in the course of the
relaxation. To gain deeper insight into the physical nature of the
suppression of the snake instability, it is instructive to perform a simple
energetic analysis. As is known, energy of a straight vortex line in BEC is
proportional to its length, and grows with the condensate density. Thus, the
vortex line oriented in the radial direction has the maximum energy in the $%
z=0$ plane, while the vertically oriented line attains the maximum energy if
it is located in the peak-density region, at distance $\rho _{0}$ from the
axis $z$ [see Eq. (\ref{potential})]. Therefore, the maximum energies of the
vertical and horizontal vortex line relate to each other as the radial, $R_{%
\text{TF}}=\sqrt{8\mu /(M\omega _{r}^{2})}$, and vertical, $Z_{\text{TF}}=%
\sqrt{8\mu /(M\omega _{z}^{2})}$, Thomas-Fermi widths of the condensate,
i.e., as aspect ratio (\ref{A})\ of the toroidal trap: $A=R_{\text{TF}}/Z_{%
\text{TF}}=\omega _{z}/\omega _{r}$. From these straightforward estimates,
it follows that the fluxons are most likely to keep the horizontal
orientation when elongated rings merge. In our numerical simulations we
observed that this property of elongated condensates does not depend on the
number of atoms, but is determined by the aspect ratio of the trap. The
robust dynamics of JVs in the elongated trap is in sharp contrast with the
evolution of fluxons in the pancake-shaped setting, which, as demonstrated
above, bend and eventually split into vertically oriented vortices and
antivortices.


The formation of $|m_{1}-m_{2}|$ fluxons at the interface between flows with
topological charges $m_{1}$ and $m_{2}$ necessarily follows from the
azimuthal periodicity of the condensate wave function. For elongated
condensates the horizontal orientation of the JVs is energetically
preferable even when the barrier between the rings vanishes, and they are
located in the peak-density regions. In the same time, vertically oriented
vortex lines remain trapped in the potential well formed by the central
toroidal hole. Eventually, this scenario gives rise to formation of
long-lived transient states in the form of the above-mentioned hybrids,
which are qualitatively similar to those supported by means of a completely
different mechanism in Ref. \cite{peanut}. The hybrids are topological
structures with $(m_{1},m_{2})$ vertically oriented vortex lines and $%
|m_{1}-m_{2}|$ horizontally oriented JVs.


Furthermore, as illustrated in Fig. \ref{fig:FixCore} (see also Supplemental
movie \cite{sup_mat}), the rotational fluxons can be pinned and completely stabilized if
the amplitude of the barrier is reduced to some nonzero value, maintaining a
residual potential well with the bottom point at $z=z_{0}$. As the result,
one obtains \textit{completely stable} stationary (rather than transient) 3D
hybrid vortex complexes with any value of the angular momentum per particle
in the range of $m_{1}<L_{p}<m_{2}$.
This value can be readily tuned by choosing position of the barrier, $z_{0}$%
, in the initial state, and even dynamically tuned by slow motion of the
sheet beam after formation of the stable hybrid as illustrated in Fig. \ref%
{fig:FixCore} for $t>0.3$ s. The stabilized JVs are not only topological
modes of fundamental interest in physics of superfluidity and
superconductivity, but also may be used for design of topologically
protected qubits, cf. Ref. \cite{qubit}.




\section{Conclusion}

We have demonstrated that the evolution of merging persistent currents in
parallel BEC\ rings is driven by essentially three-dimensional nonlinear dynamics of
JVs (Josephson vortices, alias rotational fluxons). Quite different
dynamical scenarios are reported for the fluxons, which give rise to
different final states of the condensate, depending on the imbalance in
initial populations of the two rings, and on the aspect ratio of the
toroidal trap. In sharp contrast to classical flows with viscosity, in the
considered weakly dissipative quantum system one can observe transitions
into states with higher-order values of the angular momentum. It turns out
that for elongated merging condensates the vorticity of the final state is
always determined by the more populated ring component, while for
pancake-shaped rings there is a threshold value of imbalance, such that
below it the final vorticity is zero, while above the threshold the dominant
component imparts its vorticity to the final state.

In the course of the merger of the rings carrying different initial
vorticities, which are strongly elongated in the axial direction, we
observed formation of \textit{dynamical hybrids}, which are long-lived
transient states in 3D, with different vorticities, $m_{1}$ and $m_{2}$, in
their two axially separated parts, which share a common
vertically oriented axis threading the merged double-ring system and $%
|m_{1}-m_{2}|$ radially oriented JVs (rotational fluxons). Under the action
of weak dissipation, which drives slow vertical drift of the fluxons, the
dynamical hybrid suffers slow degradation into the state with the vorticity
imposed by the initially dominant component. The JVs can be pinned and
stabilized by a residual nonvanishing potential barrier separating the
parallel rings, which thus leads to the formation of completely stable 3D
hybrid vortex complexes.

In addition to revealing the dynamics of the merger of superfluid rings with
different vorticities, and possibilities of the formation of transient and
permanent hybrids incorporating different vorticities in their axially
separated parts, these findings may also help to understand the
correspondence between rotational fluxons in long Bose-Josephson junctions
and the formation of vortex lattices at the interface of counter-propagating
superflows.


\section*{Acknowledgment}

The work of B.A.M. was partly supported by the Israel Science Foundation
through grant No. 1287/17.

\bibliography{Merging_rings_refs}

\begin{thebibliography}{48}
\bibitem [{sup()}]{sup_mat}%
  \BibitemOpen
  \href@noop {} {}\bibinfo {note} {{We present animations of dynamics of
  vortices with small and big initial imballance $P$, degradation of hybrid
  system in time and pinning possibility in systems of hybrid
  type.}}\BibitemShut {Stop}%
\bibitem [{pha()}]{phase_detect}%
  \BibitemOpen
  \href@noop {} {}\bibinfo {note} {{To detect the vortex cores we use an
  algorithm of numerical phase unwrapping at each point of the grid in
  different planes: $z=$const, $x=$const or $y=$const. For vertically oriented
  vortex lines (detected in $z=$const plane) we use red (vortices) and blue
  (antivortices) colors. For horizontally oriented cores (detected in $x=$const
  or $y=$const planes) we use black color. We note that our algorithm of phase
  unwrapping fails to detect a precise position of the vortex core in the empty
  central spatial region. Thus instead of phase unwrapping in each point for
  the central vortex line we calculate a phase shift in the plane $z=$const at
  the circle of radius $\rho=\rho_1$, where $\rho_1$ is internal TF radius of
  the condensate. The central vortex line is marked as a dashed red
  line.}}\BibitemShut {Stop}%
\end{thebibliography}

\end{document}